\documentclass[aps,prl,twocolumn,showpacs,superscriptaddress]{revtex4-1}
\usepackage{amsmath,amssymb}
\usepackage{graphicx}% Include figure files
\usepackage{dcolumn}% Align table columns on decimal point
\usepackage{bm}% bold math
\begin{document}
\title{Comment on ``Integrability of the Rabi Model''}

\author{Qing-Wei Wang}
\email[]{qwwang@theory.issp.ac.cn}
\affiliation{Department of Physics, Renmin University of China, Beijing 100872, China}
\affiliation{Key Laboratory of Materials Physics, Institute of Solid State Physics, Chinese Academy of Sciences, Hefei 230031, P. R. China}
\author{Yu-Liang Liu}
%\email[]{ylliu@ruc.edu.cn}
\affiliation{Department of Physics, Renmin University of China, Beijing 100872, China}

\date{\today}

%\begin{abstract}
% insert abstract here
%\end{abstract}
\pacs{03.65.Ge, 02.30.Ik, 42.50.Pq}
%%--03.65.Ge= Solutions of wave equations: bound states
%%--02.30.Ik= Integrable systems
%%--42.50.Pq= Cavity QED
\maketitle

%------------------------------------------------------------------------------
%-----Key point --------------------
In a recent Letter \cite{Braak2011}, Braak proved that the spectrum of the Rabi model
$H=\omega a^\dag a +g\sigma_z(a+a^\dag) +\Delta \sigma_x$
consists of regular and exceptional spectrum. The \emph{necessary and sufficient} condition for the occurrence of the exceptional eigenvalues $E_n=n\omega-g^2/\omega$ reads $K_n(n\omega)=0$, which is just the condition for Judd's solutions \cite{Judd1979,*Kus1985}.
%---------
In this Comment, we show that $K_n(n\omega)=0$ is only a \emph{sufficient} but not \emph{necessary} condition for the occurrence of $E_n$. In other words, the set of Judd's solutions is just a subset of the exceptional eigenvalues.

%-------Argument------------------------
We first solve the spectrum using Hill's determinant method \cite{wqw2013}. In Bargmann representation, the eigenvalue equation reads ($\hbar=\omega=1$)
\begin{equation}\label{spec:eigeneq}
  \left[z\partial_z +g\sigma_z (z+\partial_z)+\Delta \sigma_x \right] \psi(z) =E\psi(z),
\end{equation}
where $\psi(z)$ is an entire function of $z$. By writing $\psi(z)=e^{-gz}\sum_{m=0}^\infty (z+g)^m (p_m, q_m)^T$, inserting into Eq.(\ref{spec:eigeneq}) and eliminating $p_m$, we obtain a set of linear equations for $q_m$, $W_0^\infty q_0^\infty =0$, where $q_0^m = (q_0, q_1, \ldots,q_m)^T$ and the coefficient matrix
\begin{equation*}%\label{spec:coefficient:matrix}
  W_0^m =
   \left[
   \begin{array}{ccccc}
   a_0 & -b_0 & 0 & \cdots & \cdots \\
   -c_1 & a_1 & -b_1 &\cdots &\cdots \\
   0 & -c_2 & a_2 & \cdots &\cdots \\
   \cdots &  \cdots & \cdots & \cdots & \cdots \\
   \cdots & \cdots & 0 & -c_m & a_m \\
   \end{array}
   \right]
\end{equation*}
is tridiagonal, with $a_m=(m-x)(m-x+4g^2)-\Delta^2, b_m=2g(m+1)(m-x), c_m=2g(m-x)$, and $x= E+g^2$.
Following the analysis in Ref. \cite{wqw2013}, the eigenvalues can be determined by the equation
\begin{equation}\label{spec:eigeneq:det}
  \tilde{D}(x)\equiv \lim_{m\rightarrow \infty} \frac{\Gamma^2(m+1+x)}{\Gamma^4(m+1)} \det \left[W_{0}^m \right] =0,
\end{equation}
and the corresponding coefficients $q_0^\infty$ is the minimal solution of $W_0^\infty q_0^\infty =0$.
%In contrast to Braak's function $G_\pm(x)$, the function $\tilde{D}(x)$ is an entire function of $x$ and hence all the eigenvalues can be treated on the same ground.

\begin{figure}
  \centering
  \includegraphics[width=0.48\textwidth]{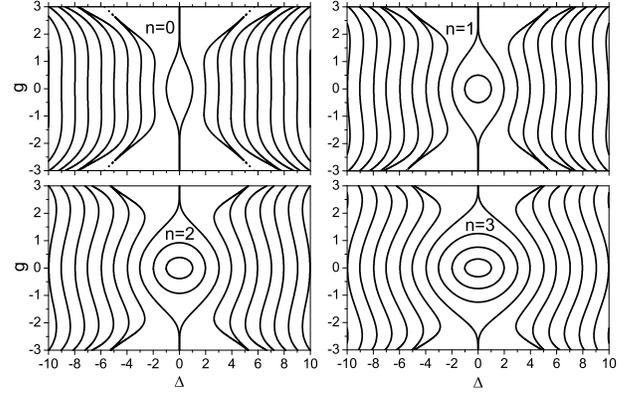}
  \caption{Solutions of $\lim_{m\rightarrow \infty} \frac{[(n+m)!]^2}{(m!)^4} \det [W_{n+1}^m]=0$ for $n=0,1,2,3$ in the $\Delta$-$g$ plane. } \label{fig:detW}
\end{figure}

Now consider the eigenvalues of the form $x_n=n$, with nonnegative integer $n$. Since $a_n=-\Delta^2, b_n=c_n=0$, we have
\begin{equation*}%\label{}
  \tilde{D}(x_n)=-\Delta^2 \det[W_0^{n-1}] \lim_{m\rightarrow \infty} \frac{[(n+m)!]^2}{(m!)^4} \det [W_{n+1}^m],
\end{equation*}
with $\det[W_0^{-1}]\equiv 1$. The eigenvalue equation (\ref{spec:eigeneq:det}) then leads to (i) $\Delta=0$, or (ii) $\det[W_0^{n-1}]=0$, or (iii) $\lim_{m\rightarrow \infty} \frac{[(n+m)!]^2}{(m!)^4} \det [W_{n+1}^m]=0$. The case (i) is the adiabatic limit and exactly solvable. The case (ii) corresponds to the isolated exact solutions and the condition $\det[W_0^{n-1}]=0$ is equivalent to $K_n(n\omega)=0$ in Braak's paper \cite{Braak2011}. The corresponding eigenvector $q_0^\infty$  has the form $(q_0,q_1, \cdots,q_{n-1}, 0,0, \cdots)^T$. The case (iii) is not discussed in Braak's paper and is the main point of this Comment. Some numerical solutions in this case are plotted in Fig.\ref{fig:detW}. We see that the solutions for a given integer $n$ contain two parts: ($\alpha$) $n$ closed loops around the center on which $\det[W_0^{n-1}]=0$ is also valid due to the double degeneracy of the corresponding level $x_n$, and ($\beta$) infinitely many lines passing through the points $g=0,\Delta=\pm(n+1),\pm(n+2),\cdots$, on which the level $x_n$ is not degenerate. These lines are neglected in Braak's solution. The corresponding eigenvector $q_0^\infty$ in case (iii) has the form $(0,0, \cdots,0, q_{n+1}, q_{n+2}, \cdots)^T$, i.e., the first $n+1$ components are zero.

%----Summary-----------------------
In conclusion, using Hill's determinant method \cite{wqw2013} we have shown that the set of Judd's solutions is only a subset of all the eigenvalues with the form $E_n=n\omega-g^2/\omega$ in the spectrum of the Rabi model. Therefore Braak's solution is not complete. We note that this structure of the exceptional spectrum has also been derived recently in Ref.\cite{Maciejewski2014}, but their method is quite different from ours.

%------Acknowledgement-----------------------------------------

%-----------Fig.---------------------------------------------

%---------bibliography---------------------------------------
%\bibliography{QRM-Braak-comment-bib}

%merlin.mbs apsrev4-1.bst 2010-07-25 4.21a (PWD, AO, DPC) hacked
%Control: key (0)
%Control: author (8) initials jnrlst
%Control: editor formatted (1) identically to author
%Control: production of article title (-1) disabled
%Control: page (0) single
%Control: year (1) truncated
%Control: production of eprint (0) enabled
%

\end{document}